\documentclass[aps,prl,reprint,superscriptaddress,longbibliography]{revtex4-1}

\usepackage[utf8]{inputenc}
\usepackage[english]{babel}
\usepackage{amsmath}
\usepackage{amssymb}
\usepackage{graphicx}
\usepackage{bm}

\DeclareMathOperator{\sech}{\mathrm{sech}}
\DeclareMathOperator{\am}{\mathrm{am}}
\DeclareMathOperator{\dn}{\mathrm{dn}}

\begin{document}
\title{Cavity Optomechanics of Topological Spin Textures in Magnetic Insulators}

\author{Igor \surname{Proskurin}}
\affiliation{Department of Physics and Astronomy, University of Manitoba, Winnipeg, Manitoba R3T 2N2 Canada}
\email{Igor.Proskurin@umanitoba.ca}
\affiliation{Institute of Natural Sciences and Mathematics, Ural Federal University, Ekaterinburg 620002, Russia}

\author{Alexander~S. \surname{Ovchinnikov}}
\affiliation{Institute of Natural Sciences and Mathematics, Ural Federal University, Ekaterinburg 620002, Russia}
\affiliation{Institute for Metal Physics, Ural Division of the Russian Academy of Sciences, Ekaterinburg 620137, Russia}

\author{Jun-ichiro \surname{Kishine}}
\affiliation{Division of Natural and Environmental Sciences, The Open University of Japan, Chiba 261-8586, Japan}

\author{Robert~L. \surname{Stamps}}
\affiliation{Department of Physics and Astronomy, University of Manitoba, Winnipeg, Manitoba R3T 2N2 Canada}

\date{\today}

\begin{abstract}
	Collective dynamics of topological magnetic textures can be thought of as a massive particle moving in a magnetic pinning potential.  We demonstrate that inside a cavity resonator this effective mechanical system can feel the electromagnetic radiation pressure from cavity photons through the magneto-optical inverse Faraday and Cotton-Mouton effects.  We estimate values for the effective parameters of the optomechanical coupling for two spin textures -- a Bloch domain wall and a chiral magnetic soliton lattice.  The soliton lattice has magnetic chirality, so that in circularly polarized light it behaves like a chiral particle with the sign of the optomechanical coupling determined by the helicity of the light and chirality of the lattice.  Most interestingly, we find a level attraction regime for the soliton lattice, which is tunable through an applied magnetic field.
\end{abstract}

\maketitle

\emph{Introduction.}---Cavity optomechanics is an established area in which the effect of radiation pressure on mechanical objects inside microwave cavity resonators is studied \cite{Aspelmeyer2014}.  The scope includes an impressive variety of phenomena including laser cooling \cite{Stenholm1986}, parametric instability \cite{Braginsky2001,Kippenberg2005} and chaotic dynamics \cite{Carmon2005}, optomechanical entanglement \cite{Palomaki2013} and nonclassical photon states \cite{Bose1997}. There are a number of applications ranging from high-accuracy sensors for gravitational wave detectors \cite{Braginsky2002,Marshall2003} to quantum information processing protocols based on the fundamental principles of the quantum mechanics \cite{Dong2012,Aspelmeyer2014}.

It was demonstrated recently, both experimentally \cite{Zhang2015,Zhang2016} and theoretically \cite{Liu2016,Kusminskiy2016}, that magnons in magnetic insulators can also feel electromagnetic radiation pressure forces owing to magneto-optical interactions such as the inverse Faraday and Cotton-Mouton effects. These findings established a new direction -- cavity \emph{optomagnonics} which has developed rapidly during the last few years \cite{Hisatomi2016,Zhang2016a,Gao2017,Sharma2017,Pantazopoulos2017,Osada2018}. On a quantum level, optomagnonics describes systems of coupled photons and magnons using an optomechanical Hamiltonian with which various optomechanical effects in magnetic insulators are predicted and described \cite{Sharma2018}. Cavity optomagnonics, together with spin optodynamics of cold atoms \cite{Brahms2010,Kohler2017}, shapes the basis of modern optospintronics \cite{Nemec2018}, which targets ultrafast optical control of spin states utilizing cavity resonators to condense the electromagnetic energy \cite{Haigh2015,Bourhill2016,Bai2017,Harder2017,Yao2017,Zhang2017,Zhang2017,Grigoryan2018,Haigh2018}.

In this paper, we propose a realization of coupled spin-photon dynamics that connects cavity optomechanics with optomagnonics.  It is well established that low-energy dynamics of modulated spin textures, such as domain walls in ferromagnets \cite{Bouzidi1990, Takagi1996,Tatara2004,Obata2008,Tatara2008,Janda2017} or magnetic soliton lattices in helimagnets \cite{Bostrem2008,Bostrem2008a,Kishine2010,Kishine2012,Kishine2016}, can be described in terms of a few collective coordinates. A canonical example is the equation of motion for a pinned domain wall (DW), which can be expressed in terms of the harmonic oscillator equation
\begin{equation}\label{DWos}
m_{\mathrm{eff}} \ddot{X} + m_{\mathrm{eff}} \Gamma_{m} \dot{X} + m_{\mathrm{eff}}\Omega_{m}^{2}X = F_{\mathrm{torque}},
\end{equation}
where $X(t)$ is the position of DW, $m_{\mathrm{eff}}$ is the effective mass  determined by the spin configuration,  $\Gamma_{m}$ is the dissipation parameter, which is proportional to the Gilbert damping, and the oscillator frequency, $\Omega_{m}$, determined by the external pinning of DW to impurities or defects \cite{Tatara2008}.  The force on the right hand side is assigned to arise from a spin torque acting upon the spin texture.  In magnetic insulators, spin torque can be of magneto-optical origin \cite{Liu2016,Kusminskiy2016}. For example, in a circularly birefringent medium the electric field $\bm{\mathfrak{E}}(\omega)$ is able to generate an effective magnetic field $\bm{B}_{\mathrm{eff}} \sim \bm{\mathfrak{E}}(\omega) \times \bm{\mathfrak{E}}(\omega)^{*}$, which is able to excite collective motion of a DW, thus creating a radiation pressure force on the right-hand side of Eq.~\eqref{DWos}.

It should be mentioned that our approach is different from those in  Refs.~\cite{Liu2016,Kusminskiy2016}, where magneto-optical coupling was applied to (macro)spin dynamics. The collective motion of modulated spin textures makes their behavior similar to actual massive particles moving in the real-space potential energy profile, which means that one can realize a variety of optomechanical applications using spin textures as effective mechanical objects, which parameters can be manipulated by applying external fields.

High tunability of spin textures in combination with low magnetic damping in such materials as iron garnets makes them suitable for applications. In what follows, we discuss how this scenario can be realized for two magnetic textures -- a Bloch DW and a periodic chiral soliton lattice (CSL), which is typical for uniaxial chiral helimagnets \cite{Togawa2012,Matsumura2017}. We demonstrate that the latter can be used for the realization of level attraction in optomechanical systems \cite{Bernier2018}, which was also demonstrated recently for cavity magnon-polaritons \cite{Harder2018}. Our approach can be further generalized to two-dimensional spin textures (e.~g. magnetic skyrmions), which has attracted attention in recent studies of the optomagnonics of a microdisk with a vortex magnetization pattern \cite{Graf2018}.

%% =============================================================================
%% =============================================================================

\begin{figure}
	\centerline{\includegraphics[width=0.375\textwidth]{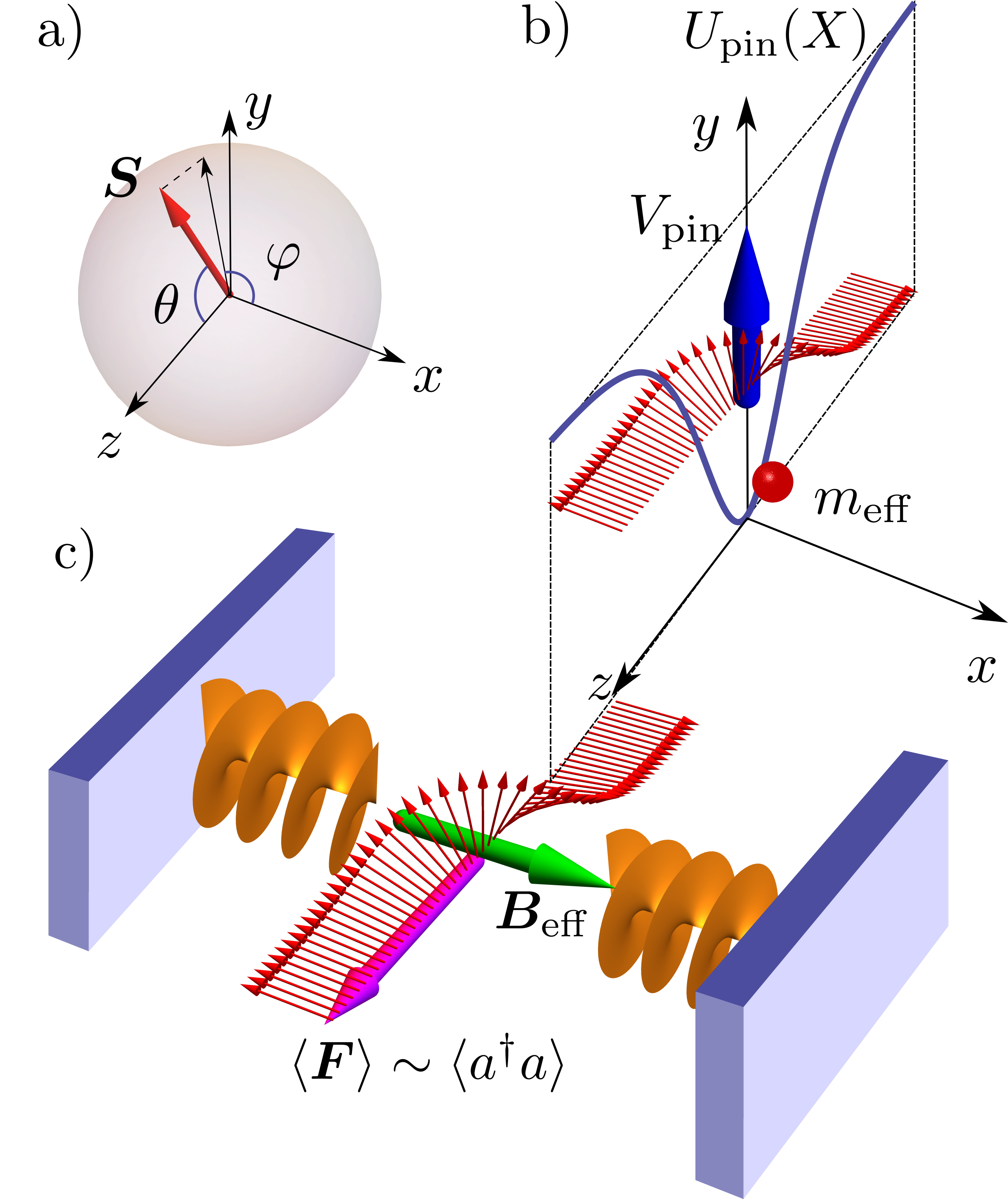}} 
	\caption{a) Coordinate system showing spin $\bm{S}(\bm{r},t)$ parametrized by the azimuthal angle $\varphi$ and the polar angle $\theta$. b) Bloch domain wall pinned at $z=0$ by the pinning interaction $V_{\mathrm{pin}}$ and the effective mechanical model: the massive particle $m_{\mathrm{eff}}$ moving in the potential energy profile $U_{\mathrm{pin}}(X)$. c) Bloch domain wall inside the cavity resonator: circularly polarized cavity mode generates the effective magnetic $\bm{B}_{\mathrm{eff}}$ through the inverse Faraday effect, which creates the optical pressure force $\langle F \rangle$ on the domain wall in the direction perpendicular to $\bm{B}_{\mathrm{eff}}$.}
	\label{fig1}
\end{figure}

\emph{The Model.}---We begin by considering a DW oscillator in the optical field. Collective dynamics of a DW can be obtained from the spin Lagrangian density \cite{Tatara2008}
\begin{equation} \label{lag}
  \mathcal{L} = \hbar S a^{-3} \left(\cos\theta - 1\right)\partial_{t}\varphi - \mathcal{H}_{M} - V_{\mathrm{pin}},
\end{equation}
which describes the semiclassical motion of the spin $\bm{S}=S(\cos\varphi\sin\theta,\sin\varphi\sin\theta,\cos\theta)$ parametrized by the angles $\varphi(z,t)$ and $\theta(z,t)$, as shown in Fig.~\ref{fig1}~(a). The second term in Eq.~\eqref{lag} is the magnetic energy density of the DW
\begin{multline} \label{Hm}
  \mathcal{H}_{M} = JS^{2}a^{-1}\left[ \left(\partial_{z}\theta\right)^{2} + \sin^{2}\theta\left(\partial_{z}\varphi\right)^{2} \right] 
  \\
  + K_{\perp}S^{2}a^{-3}\cos^{2}\theta - K_{\parallel}S^{2}a^{-3}\cos^{2}\varphi\sin^{2}\theta],
\end{multline}
where $J$ is the ferromagnetic exchange constant, $K_{\parallel}$ and $K_{\perp}$ are anisotropy parameters, and $a$ is the lattice constant. In the following, we set $S=1$ and $a=1$ and restore these factors whenever necessary. The last term in Eq.~\eqref{lag} is the pinning potential, which is modeled through a local anisotropy field at $z=0$ \footnote{Details of the pinning mechanism are not essential. Our results would remain qualitatively the same if we model the pinning, for example, via the local magnetic field or $sd$ coupling to a magnetic impurity.}, with $V_{\mathrm{pin}}=-K_{\mathrm{pin}}\delta(z)\sin^{2}\varphi\sin^{2}\theta$.

The static Bloch DW configuration, stabilized by competition of the exchange interaction with the anisotropy energy, is characterized by $\theta_{0}=\pi/2$ and $\tan\varphi_{0}(z)/2 = \exp(z/\lambda_{\mathrm{DW}})$, where $\lambda_{\mathrm{DW}} = \sqrt{J/K_{\parallel}}$ denotes the domain wall width \cite{[{See e.~g. }] Buschow2003}.

The lowest-energy dynamics of DW can be described by the collective coordinate method \cite{Bouzidi1990}. For this purpose, we introduce two dynamical variables: the position of the wall, $X(t)$, responsible for its translational motion, $\varphi(z,t) = \varphi_{0}(z - X(t))$, and the amplitude $\xi(t)$ of the out-of-plane component, $\theta(z,t) = \pi/2 +\xi(t)u_{0}(z-X(t))$. The spatial profile of $\theta(z,t)$ is determined by the P\"{o}schl-Teller equation, which gives $u_{0}(z) = \sech(z/\lambda_{\mathrm{DW}})/\sqrt{2}$ \cite{Bouzidi1990}.

In terms  of collective variables, the effective Lagrangian for the translational motion of the Bloch domain wall motion takes the following form
\begin{equation} \label{Lag}
  \mathcal{L}_{\mathrm{eff}} = \hbar \mathcal{M} \xi(t)\dot{X}(t) - \Delta \xi^{2}(t) - U_{\mathrm{pin}}(X),
\end{equation}
where $\mathcal{M}=\int u_{0}\partial_{z}\varphi_{0}dz$ and $\Delta=2K_{\perp}\lambda_{\mathrm{DW}} + \pi^{2}K_{\mathrm{pin}}/4$, and we have kept only leading order terms in $\xi(t)$. The last term in this equation is the potential energy of the pinning given by $U_{\mathrm{pin}}(X) = -K_{\mathrm{pin}} \sech^{2}(z/\lambda_{\mathrm{DW}})$. Spin relaxation in DW dynamics can be taken into account by a Rayleigh dissipation function, $\mathcal{W}=\hbar\alpha/(2Sa)\int (\partial_{t}\bm{S})^{2}dz$, where $\alpha$ is the Gilbert damping parameter \cite{Tatara2008}. 

%% =============================================================================
%% =============================================================================

Dissipative Euler-Lagrange equations for $\mathcal{L}_{\mathrm{eff}}$ can be reduced to the second order equation of motion for a massive particle moving in viscose medium in the pinning  potential $U_{\mathrm{pin}}$:
\begin{equation} \label{eom}
  m_{\mathrm{eff}}\ddot{X} + m_{\mathrm{eff}}\Gamma_{m}\dot{X} = -\frac{\partial U_{\mathrm{pin}}}{\partial X},
\end{equation}
where $m_{\mathrm{eff}}=\hbar^{2}\mathcal{M}^{2}/(2\Delta)$ is the effective mass of DW and $\Gamma_{m}=2\alpha\hbar\mathcal{K}/m_{\mathrm{eff}}$ is the effective mechanical damping, where $\mathcal{K}=\int (\partial_{z}\varphi_{0})^{2}dz$ (see Fig.~\ref{fig1}~(b)). For strong pinning, this equation reduces to Eq.~\eqref{DWos} for a damped harmonic oscillator with $\Omega_{m}^{2}=\frac{1}{2}m_{\mathrm{eff}}^{-1}\partial^{2} U_{\mathrm{pin}}/\partial z^{2}$ \cite{Tatara2008}.

The Hamiltonian formulation for the Lagrangian in Eq.~\eqref{Lag} can be found using the general formalism of Ref.~\cite{Gitman1990} that treats the canonical momentum $P_{X}=\partial \mathcal{L}_{\mathrm{eff}}/\partial \dot{Z}=\hbar \mathcal{M}\xi$ as a constraint. This allows us to exclude $\xi(t)$ and find the effective mechanical Hamiltonian for $X(t)$ \cite{Bostrem2008,*Bostrem2008a}:
\begin{equation} \label{Hmech}
  \mathcal{H}_{m} = \frac{P_{X}^{2}}{2m_{\mathrm{eff}}} + U_{\mathrm{pin}}(X).
\end{equation}
This Hamiltonian is helpful for analyzing a quantum regime of DW motion, achieved by replacing dynamic variables with operators $\hat{x} = x_{\mathrm{ZPF}}(b + b^{\dag})$ and $\hat{p}_{x}=-im_{\mathrm{eff}}\Omega_{m}x_{\mathrm{ZPF}}(b - b^{\dag})$, where $x_{\mathrm{ZPF}}=(\hbar/2m_{\mathrm{eff}}\Omega_{m})^{1/2}$, and $b$ and $b^{\dag}$ satisfy boson commutation relations. The effective damping $\Gamma_{m}$ in this case corresponds to the decoherence rate of the quantum oscillator states.

%% =============================================================================
%% =============================================================================

\emph{Magneto-optical coupling.}---The central idea of this paper is that inside a cavity resonator, a DW can feel radiation pressure forces from the electromagnetic field similar to that of a suspended mirror in standard optomechanical applications. The microscopic mechanism behind this analogy is the magneto-optical coupling between the electromagnetic field and the spin system \cite{Cottam1986}. Local spin oscillations modulate the electric permittivity tensor resulting in an interaction energy \cite{[{See, for example, }] Landau1984}
\begin{equation}\label{Hmo}
\mathcal{H}_{\mathrm{mo}} = -\frac{\varepsilon_{0}}{4} \int \delta \varepsilon_{ij}(\bm{S}) \mathfrak{E}_{i}(\bm{r},t) \mathfrak{E}^{*}_{j}(\bm{r,t}) d^{3}r,
\end{equation}
where $\mathfrak{\bm{E}}(\bm{r},t)$ denotes the complex amplitude of the electric field, $\bm{E}(\bm{r},t)=\Re[\mathfrak{\bm{E}}(\bm{r},t)\exp(-i\omega t)]$ \cite{Kalashnikova2008,Tzschaschel2017}. The electric permittivity can be expanded in a power series of the spin density, $\delta\varepsilon_{ij}(\bm{r}) = if_{ijk}S_{k}(\bm{r}) + \beta_{ijkl}S_{k}(\bm{r})S_{l}(\bm{r})+\ldots$, where the magneto-optical coupling tensors $f_{ijk}=-f_{jik}$ and $\beta_{ijkl}=\beta_{jikl}=\beta_{ijlk}=\beta_{jilk}$ are related to the Faraday and Cotton-Mouton effects respectively \cite{Cottam1986}.

Inside the cavity, the electric permittivity determines the frequency of cavity modes, $\omega_{\mathrm{cav}}(X)$, which becomes dependent on the position of the DW \footnote{In general, $\omega_{\mathrm{cav}}$ should depend on both, the position and the momentum of the wall. For a while, we ignore the momentum dependence for simplicity.}. This is similar to how a suspended mirror modulates the frequency in optomechanics \cite{Aspelmeyer2014}. Expanding $\omega_{\mathrm{cav}}(X)$ near the local minimum, $\omega_{\mathrm{cav}}(X)a^{\dag}a = [\omega_{\mathrm{cav}} + (\partial \omega_{\mathrm{cav}}/\partial X)X + \ldots]a^{\dag}a$, we obtain the magneto-optical interaction $-GX(t)a^{\dag}a$, between the DW and the cavity photons described by the $ a^{\dag} $ and $a$ operators, with microscopic details contained in the coupling parameter $ G =  -(\partial \omega_{\mathrm{cav}}/\partial X)$.

In order to illustrate the microscopic mechanism, we consider a possible experimental setup in Fig.~\ref{fig1} (c), where electromagnetic-field standing waves along the $x$-direction interact with Bloch DW along the $z$-axis. For illustration, let us consider only the inverse Faraday effect, so that  $\delta\varepsilon_{ij} = if\epsilon_{ijk}S_{k}$, where $\epsilon_{ijk}$ is the Levi-Civita symbol. In this case, the coupling energy in Eq.~\eqref{Hmo} can be expressed in terms of the spatially uniform effective magnetic field $ B^{x}_{\mathrm{eff}} \sim i\int \bm{e_{x}}\cdot(\mathfrak{\bm{E}} \times \mathfrak{\bm{E}}^{*}) dx$ applied parallel to the magnetization direction of magnetic domains connected by DW.

As is well-known \cite{Schryer1974}, the uniform magnetic field in such configuration can move DW, so that $B^{x}_{\mathrm{eff}}$ couples directly to the domain wall position. Quantizing the electric field inside the cavity, $\bm{\mathfrak{E}}(x,t) = -i \sum_{n\lambda}(\hbar\omega_{n}/\varepsilon_{0}V)^{1/2}\sin(\pi n x/L_{x})\bm{e}_{\lambda}a_{n\lambda}$, where $\omega_{n}$  are the frequencies of cavity eigenmodes, and $\bm{e}_{\lambda} = (0,\lambda,-i)/\sqrt{2}$ ($\lambda = \pm 1$) are the polarization vectors in the helicity basis, we find an explicit expression for the magneto-optical coupling between DW and the cavity photons:
\begin{equation}\label{Hm1}
\mathcal{H}_{\mathrm{mo}} = -\hbar g_{0} (b + b^{\dag})(a^{\dag}_{R}a^{\vphantom{\dag}}_{R} - a^{\dag}_{L}a^{\vphantom{\dag}}_{L}),
\end{equation}
Here $g_{0} = \frac{1}{4}fS_{\mathrm{eff}}\omega_{\mathrm{cav}}$ is the single photon coupling for the $n$th mode with $\omega_{n} = \omega_{\mathrm{cav}}$, which is related to $ G $ as $ g_{0} = Gx_{\mathrm{ZPF}} $, and  $S_{\mathrm{eff}} = x_{\mathrm{ZPF}}A_{\perp}/V$ where  $A_{\perp}$ is the cross section of the sample, and $V$ is the volume of the cavity.  The dimensionless parameter $S_{\mathrm{eff}}$ is proportional to the total number of spins involved into collective motion.  Since this number is macroscopic, $g_{0}$ can reach the same orders of magnitude as estimated for macrospin fluctuations in optomagnonics \cite{Kusminskiy2016}.

We now estimate values for the mechanical parameters of the DW oscillator coupled to the optical field. For iron garnet ferromagnetic insulators assuming $K_{\perp} \approx 0.1$~K, $\lambda_{\mathrm{DW}} \approx 100$~nm, $a \approx 1$~nm \cite{Stancil2012}. The effective mass $m_{\mathrm{eff}} \approx \hbar^{2}/(K_{\perp}\lambda_{\mathrm{DW}}a)$ is estimated as $10^{-27}$~kg; and the oscillator frequency $\Omega_{m} \approx (2K_{\mathrm{pin}}K_{\perp}a/\hbar^{2}\lambda_{\mathrm{DW}})^{1/2} \approx 10^{9}$~s$^{-1}$. For yttrium-iron-garnet, the damping parameter can be as low as $3 \times 10^{-5}$ \cite{LeCraw1962}, which gives the quality factor $Q_{m} = \Omega_{m}/\Gamma_{m}\approx \alpha^{-1}\sqrt{K_{\mathrm{pin}}a/(2K_{\perp}\lambda_{\mathrm{DW}})} \approx 10^{4}$. For the single photon coupling, we use $f = 2c\phi_{F}\sqrt{\varepsilon}/\omega_{\mathrm{cav}}$ with $\phi_{F} = 240^{\circ}$ cm$^{-1}$, and $\varepsilon = 5$ \cite{Kusminskiy2016,Stancil2012}, which gives $g_{0} = \frac{c}{2} \phi_{F} \sqrt{\varepsilon} S_{\mathrm{eff}} = 10^{5}$~s$^{-1}$ for $S_{\mathrm{eff}} = 10^{-6}$. To drive the DW oscillator, the full coupling strength should be comparable to the oscillator energy, $g_{0}\sqrt{n_{\mathrm{cav}}} \lesssim \Omega_{m} $, where $n_{\mathrm{cav}}$ is the number of coherent cavity photons. From this relation, we estimate $ n_{\mathrm{cav}} \lesssim (\Omega_{m}/g_{0})^{2} \approx 10^{8}$. The strength of the cavity electric field can be estimated as $ E \lesssim \sqrt{n_{\mathrm{cav}} \hbar \omega_{\mathrm{cav}}/(\varepsilon_{0} V)} \approx 100$ V/m for THz photons in a centimeter-sized cavity.

%% =============================================================================
%% =============================================================================

%% =============================================================================
\begin{figure}
	\centerline{\includegraphics[width=0.475\textwidth]{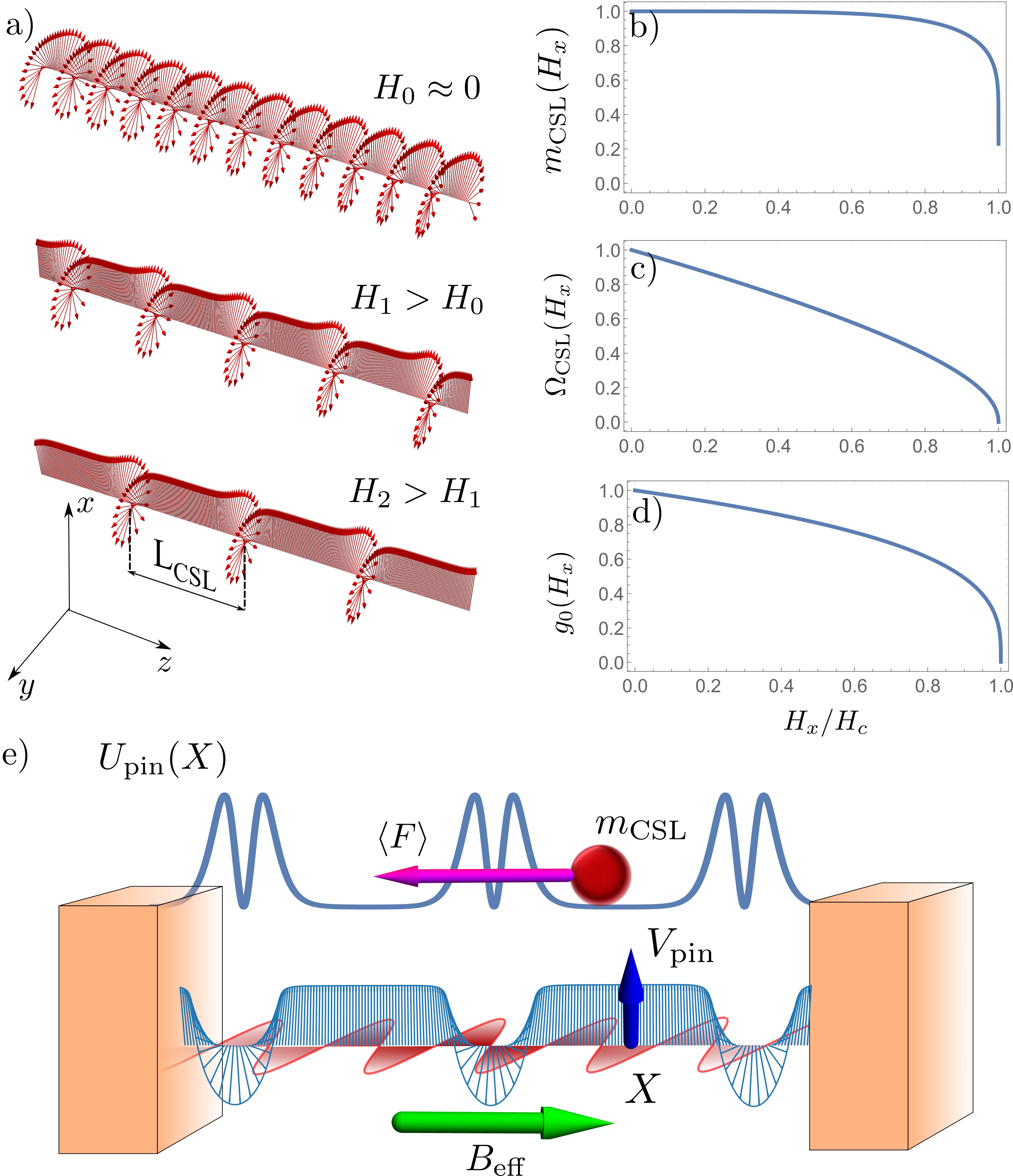}}
	\caption{CSL configuration for several $H_{x}$: $H_{0} \approx 0$, $ H_{1} > H_{0} $, and $H_{2} > H_{1}$ (a); parameters of the effective optomechanical model as functions of $H_{x}/H_{c}$ (b)--(d); CSL inside a cavity resonator: cavity modes generate $B_{\mathrm{eff}}$, which induces sliding motion of CSL, and the corresponding effective model of mass $m_{\mathrm{CSL}}$ moving inside the potential energy profile $U_{\mathrm{pin}}(X)$ determined by $V_{\mathrm{pin}}$.}
	\label{fig2}
\end{figure}
%% =============================================================================

\emph{Magnetic soliton lattice.}---Another topological structure that may have potential applications in the context of cavity optomechanics is the chiral soliton lattice (CSL), which is typically found in uniaxial chiral helimagnets \cite{Togawa2012}. Similar to a DW in that it has a twisted structure, a CSL is determined by competition between the exchange, the Dzyaloshinskii-Moriya interaction (DMI) and  the Zeeman energy in an external static magnetic field applied perpendicularly to the chiral axis. The equilibrium spin configuration of a CSL can be thought of as a periodic array of equally spaced 360$^{\circ}$ domain walls with period determined by applied magnetic field (see Fig 2 (a)). This is formally described by the Jacobi amplitude elliptic function $\varphi_{0}(z) = \pi + 2\am(2KL^{-1}_{\mathrm{CSL}}z,\varkappa)$ and is characterized by the topological winding number, $n_{\mathrm{kink}} = (2\pi)^{-1} \int \partial_{z}\varphi_{0}(z) dz $, where  $L_{\mathrm{CSL}} = 8K(\varkappa)E(\varkappa)J/(\pi D)$ is the lattice period, $K$ ($E$) is the complete elliptic integral of the first (second) kind with the elliptic modulus  $\varkappa$, and $D$ is DMI constant.  The elliptic modulus $\varkappa$ is determined by the external field $H_{x}$ via the transcendent equation, $\sqrt{H_{x}/H_{c}} = \varkappa/E(\varkappa)$, which has a solution  for $ H_{x} < H_{c} $. The critical field $2\mu_{B}H_{c}= \pi^{2} D^{2}/(16J)$ marks the incommensurate-to-commensurate transition to the forced ferromagnetic state, where $L_{\mathrm{CSL}}$ diverges and $n_{\mathrm{kink}}$ vanishes \cite{[{For review, see }] Kishine2015}.  At zero magnetic field, the CSL is reduced to a helical spin ordering.

The collective dynamics of CSL can be described in the same way as the dynamics of Bloch DW. We introduce the position $X(t)$ and the out-of-plane component $\delta\theta(z,t) = \xi(t)u_{0}(z-X(t))$, where the spatial profile $u_{0}(z) =  L_{z}^{-1/2}\sqrt{K/E}\dn(2KL_{\mathrm{CSL}}^{-1}z,\varkappa)$ is now determined by the Lam\'{e} equation. The resulting equations of motion for the CSL are identical to Eqs.~\eqref{eom} and \eqref{Hmech} with new mechanical parameters $m_{\mathrm{CSL}} = \hbar^{2} n_{\mathrm{kink}}/(a^{2}D) \mathcal{Q}_{1}^{-1}$ and $\Gamma_{m} = 2\pi^{2}\alpha D^{2}\mathcal{Q}_{1}/(\hbar J)$, where $\mathcal{Q}_{1} = \pi/(12E^{3})[(2-\varkappa^{2})E + (1-\varkappa^{2})K]$  \cite{Kishine2015}. The effective mass, $m_{\mathrm{CSL}}$, is proportional to the density of kinks, which shows a strong dependence on $H_{x}$ in the vicinity of $H_{c}$ (see Fig.~\ref{fig2} (b)). In contrast, the magnetic-field dependence of $\Gamma_{m}$  is relatively weak.

The low-energy dynamics of a pinned CSL depends on the position of the pinning site. We choose the pinning energy in the form of a local easy $x$-axis anisotropy field at the center of the ferromagnetically ordered domain, $V_{\mathrm{pin}} = -K_{\mathrm{pin}}S_{x}^{2}(z)\delta(z - \frac{1}{2}L_{\mathrm{CSL}})$ (see Fig.~\ref{fig2} (e)). In this case, in the harmonic approximation the oscillator frequency $\Omega_{\mathrm{CSL}} = (16K^{2}K_{\mathrm{pin}}L_{\mathrm{CSL}}^{-2}m_{\mathrm{CSL}}^{-1}(1-\varkappa^{2}))^{1/2}$ decreases to zero as ferromagnetically ordered regions grow with magnetic field, as shown in Fig.~\ref{fig2}~(c).

The magneto-optical coupling mechanism for CSL is different from those for Bloch DW. In order to excite collective motion of CSL, the magnetic field should be switched along the direction of the CSL axis, rather than perpendicularly to it, as for the DW, since the transverse magnetic field only modifies the period of CSL and has no impact on collective dynamics. In contrast, the magnetic field pulse applied parallel to the chiral axis couples directly to the momentum $P_{X}=\xi/(\hbar\mathcal{M})$ of the CSL and induces sliding motion \cite{Kishine2012,Kishine2016}.

To couple CSL with the optical field, we use the cavity configuration shown in Fig.~\ref{fig2} (e) where the cavity standing waves generate $\bm{B}_{\mathrm{eff}} = B_{\mathrm{eff}}^{z}\hat{\bm{z}}$ along the $z$-direction, so that $B_{\mathrm{eff}}^{z}$ couples to $S_{z} \sim \xi(t)$. Equation~\eqref{Hmo}, in this situation, gives the following coupling strength between the CSL and the cavity modes
\begin{equation}\label{Hcsl}
\mathcal{H}_{\mathrm{mo}} = i\hbar g_{0}(b - b^{\dag})(a^{\dag}_{R}a^{\vphantom{\dag}}_{R} - a^{\dag}_{L}a^{\vphantom{\dag}}_{L}),
\end{equation}
where $g_{0} = \frac{1}{4} f S_{\mathrm{eff}} \omega_{\mathrm{cav}} m_{\mathrm{CSL}} \Omega_{\mathrm{CSL}}a^{2} \mathcal{Q}_{2}/\hbar$ and $\mathcal{Q}_{2} = \mathcal{M}^{-1} \int \sin^{2}(\pi n L_{z}^{-1}z) u_{0}(z) dz$ \footnote{The derivation of Eq.~\eqref{Hcsl} is straightforward. We neglect Brillouin scattering of modes with different $n$, which is justified when cavity modes are well separated. This is exact for $H_{x}=0$ where $u_{0}(z)=\mathrm{const}$}. The coefficient $\mathcal{Q}_{2}$ does not show a strong dependence on magnetic field and can be estimated as $J/(2D)$.  In small magnetic fields, the single photon coupling strength is given by $ g_{0} = f S_{\mathrm{eff}} m_{\mathrm{CSL}} \omega_{\mathrm{cav}} \Omega_{\mathrm{CSL}}a^{2} J/(8\hbar D)$.

The sign of $g_{0}$ in Eq.~\eqref{Hcsl} is related to magnetic chirality of CSL via the sign of DMI constant $D$.  This means that in circularly polarized light, CSL behaves like a chiral mechanical particle with the sign of the radiation pressure force proportional to helicity of the light and chirality of the spin structure.

We can estimate the effective mass of CSL in zero applied magnetic field as  $m_{\mathrm{CSL}} \approx n_{\mathrm{kink}} \times 10^{-26} $~kg for $ D\approx 0.1$~K. This is approximately $n_{\mathrm{kink}}$ times larger than the mass of a single domain wall \cite{Bostrem2008a}. Typically, in millimeter-size samples $ n_{\mathrm{kink}} \approx 10^{4} $, which gives $m_{\mathrm{CSL}}\approx 10^{-22}$~kg. Taking $ K_{\mathrm{pin}} \approx 0.1 $~K and $D/J \approx 10^{-3}$, we obtain the effective mechanical frequency in zero field $\Omega_{\mathrm{CSL}} = \sqrt{\pi K_{\mathrm{pin}}D^{3}/(\hbar^{2} J^{2} n_{\mathrm{kink}})} \approx 0.1 \times 10^{6}$~s$^{-1}$, i.~e. in the megahertz range, and the effective damping $\Gamma_{m} = 2\pi\alpha D^{2}/(\hbar J)$ gives a quality factor $Q_{m} = \alpha^{-1} \sqrt{K_{\mathrm{pin}}/(4\pi n_{\mathrm{kink}} D)} \approx 10^{-2}\alpha^{-1}$ for $K_{\mathrm{pin}} \approx 10^{-3} J$. For CSL, we estimate $ g_{0}  \approx 1.2 \times 10^{6}$~s$^{-1}$ using the same optical parameters as for DW.

%% =============================================================================
%% =============================================================================
\begin{figure}
	\centerline{\includegraphics[width=0.35\textwidth]{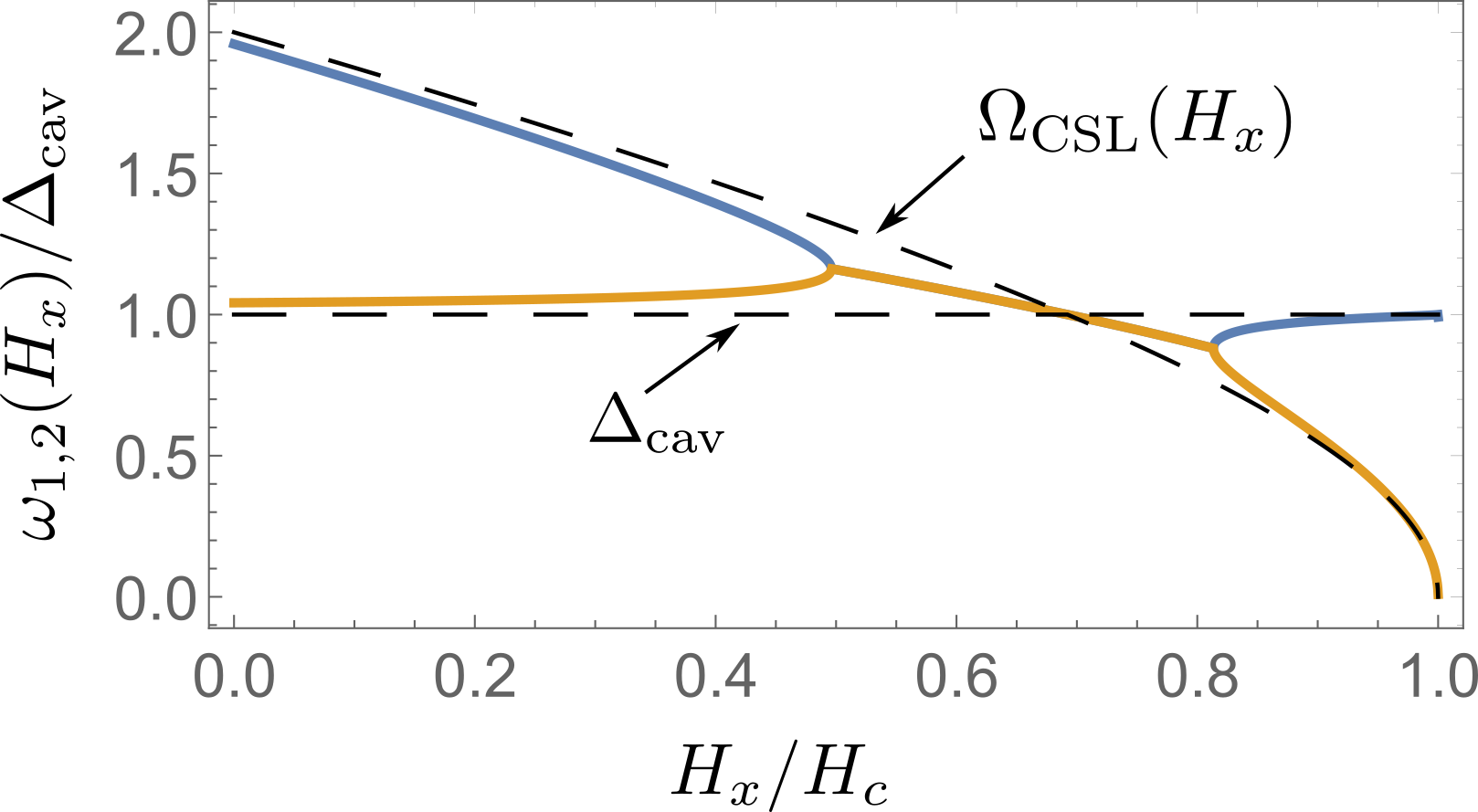}}
	\caption{Level attraction between the CSL mode $\Omega_{\mathrm{CSL}}(H_{x})$ and the cavity mode with the detuning  parameter $\Delta_{\mathrm{cav}}$ for $\Omega_{\mathrm{CSL}}(0)/\Delta_{\mathrm{cav}} = 2$ and $ g(0)/\Delta_{\mathrm{cav}} = 0.2 $. Dashed lines shows both modes without interaction.} 
	\label{fig3}
\end{figure}
%% =============================================================================
%% =============================================================================

\emph{Discussion.}---The optomechanical Hamiltonian with coupling terms in Eqs.~\eqref{Hm1} and \eqref{Hcsl} can be useful for realizing various optomechanical applications \cite{Aspelmeyer2014}, for example, such as optical cooling of the domain wall motion by analogy with optical cooling of magnons proposed recently in Ref.~\cite{Sharma2018}. 
%Low spin damping and the high tunability of such structures as a CSL are key %ingredients for functionality and applications. 

For illustration, we propose a CSL for realization of the level attraction picture, proposed recently in Ref.~\cite{Bernier2018}, using the applied static magnetic field as a control parameter to drive  the coupled system towards instability.

To realize level attraction, we use the linearized optomechanical Hamiltonian in the rotating wave approximation for a blue detuning regime \cite{Aspelmeyer2014}
\begin{equation}\label{Hcsl1}
\mathcal{H} = -\hbar\Delta_{\mathrm{cav}} a^{\dag} a + \hbar\Omega_{\mathrm{CSL}}(H_{x}) b^{\dag}b + i\hbar g(H_{x})(ab - a^{\dag}b^{\dag}),
\end{equation}
where $\Delta_{\mathrm{cav}} >0$ is the detuning parameter, $g(H_{x}) = g_{0}(H_{x})\sqrt{n_{\mathrm{cav}}}$ denotes the full optomechanical coupling strength, $ a $ and $ a^{\dag} $ denote the fluctuating part of the cavity field, such as $ a_{R} = \sqrt{n_{\mathrm{cav}}} + a $. We consider only the right-polarized mode.

The  eigenfrequencies of the Hamiltonian in Eq.~\eqref{Hcsl1} are given by \cite{Bernier2018} $\omega_{1,2}(H_{x}) = \frac{1}{2}(\Delta + \Omega_{\mathrm{CSL}}) \pm \sqrt{\frac{1}{4}(\Delta - \Omega_{\mathrm{CSL}}) - g^{2}}$. This shows level attraction in the region $2g(H_{x}) <|\Delta - \Omega_{\mathrm{CSL}}(H_{x})|$ bounded by two exceptional points where the real parts of the frequencies coalesce, as shown in Fig.~\ref{fig3}. Inside this region, an instability develops that resembles synchronization of two oscillators, where the amplitude of one mode shows exponential growth while the other is suppressed \cite{Bernier2018}.

%% =============================================================================
%% =============================================================================

\emph{Summary.}---We propose to use collective dynamics of spin textures in ferromagnetic insulators as a model of mechanical subsystems in optomechanical applications using the inverse Faraday effect as a coupling mechanism. When collective dynamics are excited by cavity electromagnetic modes, spin textures move in real space similar to actual mechanical particles. Our approach is illustrated on two topological spin structures -- the Bloch domain wall and the chiral soliton lattice. The latter is a highly tunable structure with the effective mechanical parameters that depend strongly on an applied magnetic field. This fact allows us to propose it as a realization of level attraction as proposed for microwave resonators.

%% =============================================================================
%% =============================================================================

\begin{acknowledgments}
	This work was supported by a Grant-in-Aid for Scientific Research (B) (Grant No. 17H02923) and (S) (Grant No. 25220803) from the MEXT of the Japanese Government, JSPS Bilateral Joint Research Projects (JSPS-FBR), and the JSPS Core-to-Core Program, A. Advanced Research Networks. I.P. acknowledges financial support by Ministry of Education and Science of the Russian Federation, Grant No. MK-1731.2018.2 and by Russian Foundation for Basic Research (RFBR), Grant No. 18-32-00769(mol\_a). A.S.O. acknowledges funding by the RFBR, Grant No. 17-52-50013, the Foundation for the Advancement to Theoretical Physics and Mathematics BASIS, Grant No. 17-11-107, by the Government of the Russian Federation Program 02.A03.21.0006, and by the Ministry of Education and Science of the Russian Federation, Project No. 3.2916.2017/4.6. R.L.S. acknowledges the support of the Natural Sciences and Engineering Research Council of Canada (NSERC). Cette recherche a été financée par le Conseil de recherches en sciences naturelles et en génie du Canada (CRSNG).
\end{acknowledgments}

\bibliography{cavity}

\end{document}